\documentstyle[preprint,revtex]{aps}
\large
\begin{document}
\draft
\baselineskip 0.9truecm
\begin{title}
{
Spin--Charge separation in a model of two coupled chains
}
\end{title}

\author{M. Fabrizio,}

\begin{instit}
International School for Advanced Studies,
Via Beirut 4,34014 Trieste, Italy
\end{instit}
\moreauthors{A. Parola}

\begin{instit}
Dipartimento di Fisica, Universit\'a di Milano, Milano, Italy.
\end{instit}

\begin{abstract}
A  model of interacting electrons living on two chains
coupled by a transverse hopping $t_\perp$, is solved exactly by
bosonization technique.
It is shown that $t_\perp$ does modify the shape of the Fermi surface
also in presence of interaction, although charge and spin excitations
keep different velocities $u_\rho$, $u_\sigma$.
Two different regimes occur:
at short distances, $x\ll \xi = (u_\rho - u_\sigma)/4t_\perp$,
the two chain model is not sensitive to $t_\perp$, while for larger
separation $x\gg \xi$ inter--chain hopping is relevant and generates
further singularities in the electron Green function besides
those due to spin-charge decoupling.
\vskip 1.5truecm
\noindent
PACS: 75.10.Lp, 75.10.Jm
\vskip 2.truecm
\noindent
\centerline{Ref. S.I.S.S.A. 150/92/CM/MB}
\noindent
\end{abstract}
\vskip 0.8truecm
\newpage
\narrowtext

A central problem in the theory of strongly correlated systems is
whether the mechanism leading to the breakdown of Fermi liquid (FL)
theory in one dimensional (1D) models can be generalized to higher
dimensions\cite{Anderson}.
In most 1D systems, the non FL behavior manifests itself in two distinct
ways: the correlation functions show power law behavior with
coupling dependent anomalous exponents\cite{Haldane} leading, in particular,
to the smoothing of the Fermi surface
that is often  considered as the key signature of non FL behavior.
An independent, but equally effective mechanism which induces the
breakdown of FL without generating anomalous exponents, is
the well known {\it spin--charge decoupling}\cite{Schulz} which
originates from the {\it dynamical} independence of charge and spin
excitations.
Many unsuccessful attempts have been made in the past years
to search for a breakdown of FL theory in 2D through the study
of the discontinuity in the momentum distribution, i.e. by
looking for anomalous exponents.
In this Letter we address the problem of the breakdown of FL in 2D
models\cite{PWA,Weng}
by looking directly for spin--charge decoupling. In particular, we
analyse the stability of spin--charge decoupling with respect to the
introduction of transverse hopping between two chains. We formulate
and solve via bosonization technique, a model which shows non FL behavior
without anomalous correlation exponents and we calculate several physical
quantities. At low energy we find new collective excitations
related to the transfer of an electron across the chains
which modify the analytic structure of the Green function
without restoring FL behavior.
Some results of the two chain problem can  be easily extended to
an array of $N$--chains, i.e to  strongly anisotropic 2D systems.

A simple 1D model which shows spin--charge decoupling without
anomalous power--law decay in the correlation functions, can be obtained
by including only forward scattering processes between electrons on the
same branch of the Fermi surface (usually referred to as $g_4$ interactions).
The hamiltonian of this model is\cite{Solyom}:
\begin{eqnarray}
\hat{H}_0&=&\sum_{k,\sigma} v_F(k-k_F)a^\dagger_{k\sigma}
a^{\phantom{\dagger}}_{k\sigma}+
\sum_{k,\sigma} v_F(-k-k_F)b^\dagger_{k\sigma}
b^{\phantom{\dagger}}_{k\sigma} +\nonumber\\
&+& \frac{1}{2L}\sum_{k_1,k_2,p}\sum_{\alpha,\beta}
(g_{4||}\delta_{\alpha\beta}+g_{4\perp}\delta_{\alpha-\beta})
(a^\dagger_{k_1\alpha}a^\dagger_{k_2\beta}
a^{\phantom{\dagger}}_{k_2+p\beta}a^{\phantom{\dagger}}_{k_1-p\alpha}+
b^\dagger_{k_1\alpha}b^\dagger_{k_2\beta}
b^{\phantom{\dagger}}_{k_2+p\beta}b^{\phantom{\dagger}}_{k_1-p\alpha})
\nonumber \\
\label{ham}
\end{eqnarray}
where the operator $a^\dagger_{k\sigma}$($b^\dagger_{k\sigma}$) creates
an electron of momentum $k$ and spin $\sigma$ belonging to the branch with
positive(negative) slope $v_F$($-v_F$).
Due to the linear dispersion relation of the hamiltonian\cite{Luttinger}
(\ref{ham}),
this model can be solved by use of standard bosonization technique
being a special case of the well known Tomonaga--Luttinger model.
While this interaction does not affect the
ground state properties,
the excitations are profoundly different. Two low--lying gapless
excitations with linear dispersion, are found,
describing charge and spin collective modes with different sound
velocities $u_{\rho}=v_F+(g_{4||}+g_{4\perp})/2\pi$ and
$u_{\sigma}=v_F+(g_{4||}-g_{4\perp})/2\pi$.
No quasi--particle
excitation exists and as a consequence the Green function does not
have simple poles but branch cut singularities.
The bosonization method
allows to calculate the exact form of the Green function in real space
and time:
\begin{eqnarray}
G(x,t)&=&\frac{1}{2\pi}\left\{
\frac{ {\rm e}^{ {\rm i}k_Fx } }{
\sqrt{ (x-u_\rho t + {\rm i}\eta{\rm sign}t)
(x-u_\sigma t + {\rm i}\eta{\rm sign}t) } } +\right. \nonumber\\
&-&\left. \frac{ {\rm e}^{ -{\rm i}k_Fx } }{
\sqrt{(x+u_\rho t - {\rm i}\eta{\rm sign}t)
(x+u_\sigma t - {\rm i}\eta{\rm sign}t) } }\right\}.
\label{green}
\end{eqnarray}
After Fourier transforming, the spectral weight, e.g. for the
right moving electrons, can be easily obtained (for more general
1D models see Ref. 7):
\begin{eqnarray}
A(k_F+q,\omega)&=&\frac{\left[
\theta(q)\theta(\omega-u_\sigma q)\theta(u_\rho q - \omega) +
\theta(-q)\theta(\omega-u_\rho q)\theta(u_\sigma q -\omega) \right]}
{\pi\,\sqrt{|\omega-u_\sigma q ||\omega-u_\rho q |} }.
\label{spectral}
\end{eqnarray}
In the particular case $g_{4||}=g_{4\perp}=0$ the spin and charge
velocities $u_\sigma$ and $u_\rho$ coincide and the branch cut
merges in a simple pole reproducing the standard free particle Green
function.
In this simple
model, although the low lying excitations can not be described
in terms of the Landau Fermi liquid theory, all the
equal time correlation functions at zero temperature,
including the momentum distribution,
coincide with the non--interacting ones.
The form of the spectral function (\ref{spectral})
explicitly shows that non FL behavior can occur even without anomalous
exponents. In this case, the breakdown of FL is related to
a change in the analytic structure of $A(k,\omega)$ which
does not generate the logarithmic singularities in perturbation theory
present in many 1D models.

This model is a prototype to study the relevance of spin--charge
decoupling in a system of chains coupled by transverse hopping.
As a first step let us consider the simple case of two chains.
The total hamiltonian will be the sum of two terms like Eq. (\ref{ham}),
one for each chain, plus an hopping term between the two chains:
\begin{equation}
H_\perp=
-t_\perp \sum_{k\sigma} \left(
a^\dagger_{k\sigma,1}a^{\phantom{\dagger}}_{k\sigma,2}+
b^\dagger_{k\sigma,1}b^{\phantom{\dagger}}_{k\sigma,2}+
{\rm H.c.}
\right)
\label{tperp}
\end{equation}
where the suffixes 1 or 2 refer to the two chains.
We will now show how it is possible to solve exactly this problem
by bosonization technique. As there is no term in the Hamiltonian which
couples right to left moving electrons, we treat only the case
of the right moving electrons. The hamiltonian Eq. (\ref{ham}) can be easily
rewritten in terms of the density operators of the electrons on the positive
branch\cite{Solyom}:
\begin{eqnarray}
\hat{H}_0&=& \frac{2\pi}{L} u_\rho \sum_{q>0} \left[
\rho_1(q)\rho_1(-q) + \rho_2(q)\rho_2(-q) \right] +\nonumber\\
&+& \frac{2\pi}{L} u_\sigma \sum_{q>0} \left[
\sigma_1(q)\sigma_1(-q) + \sigma_2(q)\sigma_2(-q) \right]
\label{hambos}
\end{eqnarray}
where $\rho_1(q)(\sigma_1(q))$ is the charge(spin) density operator
of the right moving electrons on chain 1. In order to write the transverse
hopping Eq. (\ref{tperp}) in terms of the density operators,
it is necessary to introduce the boson representation of fermion
operators\cite{bosferm}. Then one obtains:
\begin{equation}
H_\perp= 2{\rm i}t_\perp\frac{1}{2\pi \alpha}\int {\rm d}x
\left[ \sin{(\phi_{\rho s}(x)+\phi_{\sigma s}(x))} +
 \sin{ (\phi_{\rho s}(x)-\phi_{\sigma s}(x)) } \right]
\label{perpbos}
\end{equation}
where the operator $\phi_i$'s are the phase fields related to
the densities:
\begin{equation}
\begin{array}{lll}
\rho_c=\frac{1}{\sqrt{2}}(\rho_1+\rho_2) &~~~&
\rho_s=\frac{1}{\sqrt{2}}(\rho_1-\rho_2)\nonumber \\
\sigma_c=\frac{1}{\sqrt{2}}(\sigma_1+\sigma_2) &~~~&
\sigma_s=\frac{1}{\sqrt{2}}(\sigma_1-\sigma_2)
\end{array}
\label{densita}
\end{equation}
and they are defined as\cite{bosferm,Solyom}:
\[
\phi_i(x)={\rm i} \sum_{q>0}\frac{2\pi}{qL} {\rm e}^{-\alpha q/2}
\left( {\rm e}^{-{\rm i}qx}\rho_i(q)-{\rm e}^{{\rm i}qx}
\rho_i(-q)\right).
\]
The suffix $i$ labels any of the four densities in Eq. (\ref{densita}),
and the factor exp$(-\alpha q/2)$ plays the role of an ultraviolet
cutoff for
divergent integrals. In terms of the densities (\ref{densita}),
the hamiltonian (\ref{hambos}) remains diagonal.
Let us introduce the fermionic fields corresponding to the densities
(\ref{densita}). Then, the hopping term (\ref{perpbos}) acquires
a simple form:
\begin{equation}
-t_\perp\int {\rm d}x \left(
\Psi^\dagger_{\rho s}(x)\Psi^{\phantom{\dagger}}_{\sigma s}(x) +
{\rm H.c.}\right)+\left( {\rm e}^{2{\rm i}k_F x}
\Psi^\dagger_{\rho s}(x)\Psi^{\dagger}_{\sigma s}(x) +
{\rm H.c.}\right).
\label{hopp}
\end{equation}
By defining the Fourier transform $c_\rho(k)$ ($c_\sigma(k)$)
of the field operator $\Psi_{\rho s}$ ($\Psi_{\sigma s}$),
the total hamiltonian, i.e. Eqs. (\ref{hambos}) plus (\ref{hopp}),
can be rewritten as:
\begin{eqnarray}
\hat{H}&=&\sum_k \left(
u_\rho(k-k_F)c^\dagger_\rho(k)c^{\phantom{\dagger}}_\rho(k)+
u_\sigma(k-k_F)c^\dagger_\sigma(k)c^{\phantom{\dagger}}_\sigma(k)\right)+
\nonumber\\
&-&t_\perp \sum_k \left( c^\dagger_\rho(k)c^{\phantom{\dagger}}_\sigma(k)
+ {\rm H.c.}\right) +\nonumber\\
&-& t_\perp \sum_{k>0}\left(
c^\dagger_\rho(k+k_F)c^\dagger_\sigma(k-k_F) + {\rm H.c.} +
c^\dagger_\rho(k-k_F)c^\dagger_\sigma(k+k_F) + {\rm H.c.} \right).
\label{hamiltoniana}
\end{eqnarray}
This hamiltonian is the sum of bilinear terms of fermion operators,
therefore it can be easily diagonalized. By performing a particle--hole
transformation and then a unitary transformation,
four excitation branches are obtained:
\begin{equation}
\begin{array}{l}
\epsilon_1(q)= u_\rho q\\
\epsilon_2(q)= u_\sigma q\\
\epsilon_3(q)= \frac{1}{2}(u_\rho+u_\sigma)q+
\sqrt{\left(\frac{1}{2}(u_\rho-u_\sigma)q\right)^2+4t_\perp^2}\\
\epsilon_4(q)= \frac{1}{2}(u_\rho+u_\sigma)q-
\sqrt{\left(\frac{1}{2}(u_\rho-u_\sigma)q\right)^2+4t_\perp^2}
\label{epsilon}
\end{array}
\end{equation}
where $q=k-k_F$ is positive.
Having set the chemical potential equal to zero,
the ground state is obtained by filling the branch $\epsilon_4(q)$
up to a momentum $Q=2t_\perp/\sqrt{u_\rho u_\sigma}$
corresponding to $\epsilon_4(Q)=0$.

We have checked that, although this spinless fermion excitation spectrum
seems quite complicated, it correctly reproduces all known results,
in particular the charge-- and spin--density
normal modes in the trivial limits $a)$ $g_{4||}=0=g_{4\perp}$ for
$t_\perp\not=0$; $b)$ $g_{4||}\not=0$, $g_{4\perp}\not=0$ for
$t_\perp=0$.

{}From the ground state and the excitation spectrum
several physical properties of the system can be calculated.
The ground state energy is given by twice
the integral of $\epsilon_4(q)$ from $q=0$ up to $q=Q$.
This accounts for both right and left moving electrons, giving:
\begin{equation}
\frac{\Delta E}{L}= -\frac{1}{2\pi} \frac{4 t_\perp^2}{(u_\rho-u_\sigma)}
\log(\frac{u_\rho}{u_\sigma}).
\label{energia}
\end{equation}
Notice that Eq. (\ref{energia}) reproduces the correct result
in the non--interacting case, i.e. when $u_\rho=u_\sigma=v_F$.
The energy correction (\ref{energia})
being proportional to $t_\perp^2$ implies
a non zero value of the transverse hopping operator averaged on the
ground state, which in turns means that
spin--charge decoupling is not sufficient to generate confinement
of the electrons within each chain\cite{PWA}.
In order to analyse this issue more deeply,
we calculate the difference of the occupation numbers
between the bonding and the anti--bonding band. By labelling the two
bands according to the correspondent transverse momenta, i.e.
$k_\perp=0$ for the bonding combination, and $k_\perp=\pi$ for
the anti--bonding one, we get:
\begin{equation}
\frac{\langle N_0 - N_\pi\rangle}{L}= \frac{4t_\perp}{2\pi}
\label{nzeromenonpi}
\frac{1}{u_\rho-u_\sigma}\log\left(\frac{u_\rho}{u_\sigma}\right)
\end{equation}
which, again, does not show confinement in the two chain problem. As a further
probe for confinement, we have evaluated the number fluctuations between the
two chains. This quantity is easily related to the long wavelength limit of the
density structure factor, which can be calculated by the bosonization method:
\begin{equation}
\frac{1}{2} \frac{ \langle (N_1 -N_2)^2 \rangle }{L} =
\frac{ t_\perp }{ \pi (u_\rho-u_\sigma) }
\left( 1 - \sqrt{\frac{u_\sigma}{u_\rho}} \right).
\end{equation}
This expression, apart from the trivial dependence upon $t_\perp$, is a
function of the interaction through $u_\rho$ and $u_\sigma$.
In the simple case of spin--isotropic interaction, i.e. $g_{4||}=g_{4\perp}$,
the number fluctuation is a monotonic decreasing function of the interaction,
which goes to zero only when the coupling tends to infinity.

It is now interesting to investigate the stability of
spin--charge decoupling
with respect to the introduction of transverse hopping.
{}From a simple inspection of the
excitation spectrum in Eq. (\ref{epsilon}), one can argue that for large
momentum the simple 1D picture is recovered, together with
spin--charge decoupling. In fact, for large $q$,
$\epsilon_3(q)\to \epsilon_1(q)=u_\rho q$, $\epsilon_4(q)\to
\epsilon_2(q)=u_\sigma q$, and the spectrum tends to the $t_\perp=0$ case.
On the other hand, new excitations appear  at low energy,
centered around $q=2t_\perp/\sqrt{u_\rho u_\sigma}$, with an
almost linear spectrum and velocity $2 u_\rho u_\sigma/(u_\rho+u_\sigma)$.
The length scale which separates the two regimes is given by
\begin{equation}
\xi = (u_\rho - u_\sigma)/4t_\perp.
\label{scala}
\end{equation}

This form of the excitation spectrum leads to important consequences in
the spin and charge density--density correlation functions.
On chain 1 the latter is defined as
\[
-{\rm i}\theta(t)\langle[\rho_1(q,t),\rho_1(-q)]\rangle =
-\frac{{\rm i}\theta(t)}{2} \left\{
\langle[\rho_c(q,t),\rho_c(-q)]\rangle +
\langle[\rho_s(q,t),\rho_s(-q)]\rangle\right\}
\]
The first term on the right hand side is unaffected by $t_\perp$
and it contributes to the spectral weight ${\sl A}(\omega,q)$
with a delta--function centered at $\omega=-u_\rho q$.
For $q\xi \gg 1$, the $t_\perp\sim 0$ regime is recovered
and most of the spectral weight lies in the sharp peak at
$\omega\simeq-u_\rho q$ as in the case of two
independent chains (Fig. 1). In the opposite limit $q\xi\ll 1$
the effect of transverse hopping dominates over interaction and the
system behaves like in the non interacting case where the $\rho_s$ term
contributes to the spectral weight with
two delta--peaks centered respectively at $\omega=-v_F q +2t_\perp$
and at $\omega=-v_F q - 2t_\perp$ (see Fig. 2). In both limits, the
interplay between the transverse hopping and the interaction, has the effect
of broadening the peaks in $A(k,\omega)$,
which now acquire a finite width.

A direct probe of spin--charge decoupling can be obtained through
the single--particle Green function. This property is difficult to extract
from our solution, because the fermionic operators have a
very complicated expression in terms of the normal modes which diagonalize
the hamiltonian. However, it is possible to obtain its asymptotic
behavior at large distance which contains the relevant information
regarding FL behavior.
In this regime, we approximate the excitation
branch $\epsilon_4(q)$ in Eq. (\ref{epsilon}) near $q=Q$
with a linear spectrum and velocity $u_r=2 u_\rho u_\sigma/(u_\rho+u_\sigma)$.
This simplification allows to calculate the real space Green function
at fixed transverse momentum ($k_\perp=0,\,\pi$):
\begin{equation}
\langle \Psi_0(x,t) \Psi^\dagger_0 (0,0) \rangle \sim
{\rm e}^{{\rm i}(k_F+\Delta k_F)x} (x-u_\rho t)^{-\frac{3}{8}}
(x-u_\sigma t)^{-\frac{3}{8}}(x-u_r t)^{-\frac{1}{4}}
\label{Gdizero}
\end{equation}
where
\begin{equation}
\Delta k_F = \frac{t_\perp}{u_\rho-u_\sigma}
\log{(u_\rho/u_\sigma)}.
\label{deltakf}
\end{equation}
An analogous expression can be derived for the anti--bonding combination
of the two chains ($k_\perp=\pi$) by reversing the sign of $\Delta k_F$.
Equation (\ref{deltakf}) shows that
the transverse hopping does move the Fermi points of the bonding with respect
to the anti--bonding band.
The asymptotic expression Eq. (\ref{Gdizero}) corresponds
to the region $x, v_F t\gg \xi $; for smaller separation
we expect to recover the single chain limit Eq. (\ref{green}).
Therefore,  we find a crossover between two regimes.
On a scale $x < \xi$ fluctuations between the two chains are suppressed
and the model essentially behaves like two independent chains. At long distance
($x > \xi$), the coupling between the chains is reintroduced but interaction
still plays a crucial role. An electron excitation now decays into
a {\sl triplet} of elementary excitations, and not just a holon and a spinon.

Encouraged by the absence of singularities in the transverse hopping
as it comes out from the exact solution, we checked the bosonization
results against perturbation theory in $t_\perp$.
The first order correction to the single particle Green function
can be easily obtained up to first order in $t_\perp$:
\begin{eqnarray}
G_{0(\pi)}(k,\omega)&=& G^{(0)}(k,\omega)
\mp t_\perp \left [G^{(0)}(k,\omega)\right ]^2
\label{greenfunction}
\end{eqnarray}
where the $-(+)$ corresponds to $k_\perp=0(\pi)$, and
the unperturbed Green function $G^{(0)}(k,\omega)$ is just
the Fourier transform of Eq. (\ref{green}).
{}From  Eq. (\ref{greenfunction}) we can calculate the correction to
the bare momentum distribution $n^{(0)}_0(k)=n^{(0)}_\pi(k)=
\theta(k_F-k)$:
\[
\delta n_0(k)=-\delta n_\pi (k)= {\rm i}t_\perp\int \frac{{\rm d}\omega}
{2\pi} {\rm e}^{{\rm i}\omega 0^+} \left [G^{(0)}(k,\omega)\right ]^2.
\]
This integral gives apparently zero  due to a well known
anomaly of the $T=0$ perturbation theory\cite{Kohn}.
This problem can be avoided by working at finite temperature.
In this way the momentum distribution becomes:
\[
\delta n_0(k_F+q)= t_\perp \frac{1}{(u_\rho-u_\sigma)q}
\left[ f(\beta u_\sigma q) - f(\beta u_\rho q) \right]
\]
where the function $f(x)$ is the Fermi distribution.
By taking the $T\to 0$ limit, we get:
\[
\delta n_0(k)= \frac{t_\perp}{u_\rho-u_\sigma}
\log{(u_\rho/u_\sigma)} \delta(k-k_F)
\]
which can be interpreted as a shift of Fermi momenta given by
Eq. (\ref{deltakf}).
Analogously,
we can easily check that the total energy correction induced by this
shift coincides with Eq. (\ref{energia}).
Therefore, at least up to lowest order in $t_\perp$, perturbation
theory agrees with the exact solution.

It is straightforward to generalize these perturbative results to the case
of an array of $N$ chains. The relevant equations remain unchanged, the only
difference being that the inter--chain hopping operator acting at transverse
momentum $k_\perp$ is formally replaced by $-t_\perp \cos(k_\perp)$. For
example the correction to the longitudinal Fermi momentum with transverse
momentum $k_\perp$ is:
\[
\Delta k_F(k_\perp)= \frac{t_\perp \cos(k_\perp)}{u_\rho-u_\sigma}
\log{(u_\rho/u_\sigma)}.
\]
to first order in $t_\perp$. This equation gives the shape of the Fermi
surface for a strongly anisotropic 2D system as a function of the
spin and charge velocities. However,
we have not been able to find the exact solution of the $N$-chain problem
to all orders in $t_\perp$
and the crucial issue of the breakdown of FL behavior in two
dimensions is still open.

We thank E. Tosatti, J. Voit, and P. Nozi\`eres for helpful discussions.

\newpage

\figure{ Charge density--density spectral function in the channel
$\rho_s=1/\sqrt{2}(\rho_1-\rho_2)$. The calculation has been performed
setting $u_\rho - u_\sigma = 0.1$, $t_\perp=0.01$ and
$q=0.7$ in units of $v_F=1$.
\label{Figuno}}
\figure{ The same as Fig. 1 but at momentum $q=0.05$.
\label{Figdue}}

\bigskip
\end{document}